\documentclass[a4paper,fleqn,usenatbib]{mnras}

\usepackage[T1]{fontenc}
\usepackage{ae,aecompl}
\usepackage{booktabs}
\usepackage{pdflscape}
\usepackage{graphicx}	
\usepackage{amsmath}	
\usepackage{amssymb}	

\title[Leo A]{Chemistry in the dIrr galaxy Leo A\thanks{Based on data obtained at the Gran Telescopio Canarias.}}

\author[Ruiz-Escobedo et al.]{
Francisco Ruiz-Escobedo,$^{1}$\thanks{E-mail: fdruiz@astro.unam.mx}
Miriam Pe\~na,$^{1}$
Liliana Hern\'andez-Mart{\'\i}nez$^{2}$ and
\newauthor  Jorge Garc{\'\i}a-Rojas$^{3,4}$
\\
$^{1}$Instituto de Astronom{\'\i}a, Universidad Nacional Aut\'onoma de M\'exico, Apdo. Postal 70264,  Cd. de M\'exico, 04510, M\'exico \\
$^{2}$ Instituto de Ciencias Nucleares, UNAM\\
$^{3}$Instituto de Astrof{\'\i}sica de Canarias (IAC), E-38200 La Laguna, Tenerife, Spain\\
$^{4}$Universidad de La Laguna, Dept. Astrof\'{\i}sica. E-38206 La Laguna, Tenerife, Spain 
}

\date{Accepted XXX. Received YYY; in original form ZZZ}

\pubyear{2018}

\begin{document}
\label{firstpage}
\pagerange{\pageref{firstpage}--\pageref{lastpage}}
\maketitle

\begin{abstract}
We present chemical abundance determinations of two \ion{H}{II} regions in the dIrr galaxy Leo A, from GTC OSIRIS long-slit spectra. Both \ion{H}{II} regions are of low excitation and seem to be ionised by stars later than O8V spectral type. 
In one of the \ion{H}{II} regions we used the direct method: O$^{+2}$ ionic abundance was calculated using an electronic temperature determined from the [\ion{O}{III}] $\lambda\lambda$4363/5007 line ratio; ionic abundances of O$^+$, N$^+$, and S$^+$ were calculated using a temperature derived from a parameterised formula. O, N and S total abundances were calculated using Ionisation Correction Factors from the literature for each element. Chemical abundances using strong-line methods were also determined, with similar results. 
For the second \ion{H}{II} region, no electron temperature was determined thus the direct method cannot be used.  
We computed photoionisation structure models for both \ion{H}{II} regions in order to determine their chemical composition from the best-fitted models. It is confirmed that Leo A in a very low metallicity galaxy, with 12+log(O/H)=7.4$\pm$0.2, log(N/O)=$-$1.6, and log(S/O)=$-$1.1.
Emission lines of the only PN detected in Leo A were reanalysed and a photoionisation model was computed. This PN shows 12+log(O/H) very similar to the ones of the \ion{H}{II} regions and a low N abundance, although its log(N/O) ratio is much larger than the values of the \ion{H}{ii} regions. Its central star seems to have had an initial mass lower than 2 M$_\odot$.

\end{abstract}





\section{Introduction}
In the nearby Universe, the most common galaxies are the dwarf ellipticals and irregulars. The analysis of these small galaxies allow us to study  the history of galaxy formation and the amount and distribution of dark matter, as they seem to be dominated by it.  Observations and analysis of the chemistry in the interstellar medium (ISM) in these galaxies may help us to understand their star formation and evolution \citep[and references therein]{Hernandez-MartinezPena2009}. The study of \ion{H}{ii} regions and planetary nebulae (PNe) in irregular galaxies provides important information on the ISM metallicity.  By the analysis of the chemical abundances in both types of nebulae  we can infer the evolution of the interstellar metallicity and hence the chemical evolution of the galaxy. This can be done owing to the chemical abundances obtained from PNe (produced by low and intermediate mass stars with initial mass M $<$ 8 M$_\odot$) represent the abundance of the progenitor cloud through the elements not modified by stellar nucleosynthesis, whereas the chemical abundance of \ion{H}{II} regions  (associated to massive stars with M $>$ 8 M$_\odot$) represent the present-day chemical content of the ISM. 
    
  Leo A,  with  coordinates $\alpha$(J2000.0) = 09:59:24.8, $\delta$(J2000.0) = +30:44:57, is an isolated dIrr galaxy in the Local Group,  located at about 800 kpc from the Milky Way and at 1200 kpc from M31, and it presents a radial velocity of 22.3 km s$^{-1}$ \citep{DolphinSaha2002,McConnachie2012}. 
  Leo A is a low mass galaxy  with M$_{dyn}$ $\leq$ 2.5$\times$10$^7$ M$_\odot$, M$_{stars}$= 6$\times$10$^6$ M$_{\odot}$ and M$_{HI}$=1.1$\times $10$^7$ M$_{\odot}$ \citep[and references therein]{McConnachie2012} and shows recent star formation. Its stellar population contains old and young population components, although it conspicuously lacks a prominent ancient population, older than 10 Gyr \citep{TolstoyGallagher1998}.

  \citet{ColeSkillman2007} derived a star formation history (SFH) for Leo A founding that its first 5 Gyr were of quiescence,  most of the star formation happened more recently than 6 Gyr ago and the fraction of ancient stars was conclusively determined to be less than 10\%, although there is a small number of RR Lyr \citep{DolphinSaha2002,BernardMonelli2013}.

Some \ion{H}{II} regions were reported by \citet{StrobelHodge1991} and one PN was discovered by \citet{SkillmanKennicutt1989} from imaging later published by \citet{StrobelHodge1991}. From  spectroscopic data obtained with the 5-m Palomar telescope \citet{vanZeeSkillman2006} derived chemical abundances of four diffuse \ion{H}{II} regions and the PN.  In spite of their observational effort  (2 hrs of exposure time for the deeper spectra) these authors could not detect the auroral lines [\ion{O}{III}]  $\lambda$4363 nor [\ion{N}{ii}] $\lambda$5755, in the \ion{H}{II} regions; these lines are needed to compute the electron temperature of the plasma and, hence, they  only obtained some abundance ratios by using semi-empirical methods.  They derived an average oxygen abundance 12+log(O/H)=7.38$\pm$0.10  for the \ion{H}{II} regions. For the PN, they  measured the [\ion{O}{III}]  electron temperature and obtained 12 + log(O/H) = 7.30$\pm$0.05, by just adding the O$^+$/H$^+$ and the O$^{+2}$/H$^+$ ionic abundances. 

Owing  to its low metallicity, Leo A is an ideal laboratory to study the nature of the metal poor ISM. The analysis of chemical abundances in \ion{H}{II} regions and PNe in metal poor environments can give hints about stellar evolution (whether stellar 3rd dredge-up episode is present or not), if dIrr galaxies are chemically homogeneous, even it is possible to explore the primordial helium abundance. 

A well constrained 'one-zone' chemical evolution model (CEM) for Leo A  has been performed by \citet[][submitted, hereinafter H-M2018]{Hernandez-MartinezCarigi2018}, aiming to analyse its formation and evolution in a cosmological context and exploring the possibility of outflows driven by star formation. The main observational constraints that they included in their models are the star formation rate and the metallicity evolution obtained by \citet{GallartMonelli2015} which is part of the Local Cosmology from Isolated Dwarfs project Group (LCID)\footnote{http:\\www.iac.es/proyecto/LCID}. In addition, as the ISM abundances are very important to constrain the chemical evolution of the galaxy, they included the oxygen abundances measured by \citet{vanZeeSkillman2006} for the \ion{H}{II} regions and for the PN.

\defcitealias{Hernandez-MartinezCarigi2018}{H-M2018} 

\citetalias{Hernandez-MartinezCarigi2018} were able to reproduce the metallicity evolution and, within uncertainties, the past component of the gas for the PNe. However the O abundances predicted  by their CEM are below the observational value for the \ion{H}{II} regions (present component). They argue that the chemistry of the four \ion{H}{II} regions analised by \citet{vanZeeSkillman2006} were computed by semi-empirical and strong-line methods and their uncertainties are quite large, so it is necessary to have better  determinations.  

Our main goal in this work is to obtain accurate chemical abundances of two \ion{H}{II} regions in Leo A.  The observations, obtained with the 10.4-m Gran Telescopio Canarias (GTC), and data reduction are described in \S 2. Line intensities, physical conditions and ionic abundances  are presented in \S 3, where we have included the observations for the PN as given by \citet{vanZeeSkillman2006} to calculate its physical conditions using the same methodology as for \ion{H}{II} regions. In \S 4  the total abundances derived for the \ion{H}{II} regions and the PN, from different methods (direct method and strong line methods) are shown. Photoionisation structure models computed with the code CLOUDY for the observed \ion{H}{II} regions and the PN are described in \S 5  and our results are discussed in \S 6.

\section{Observations and data reduction}

\begin{figure*} 
\includegraphics[width=1.0\columnwidth]{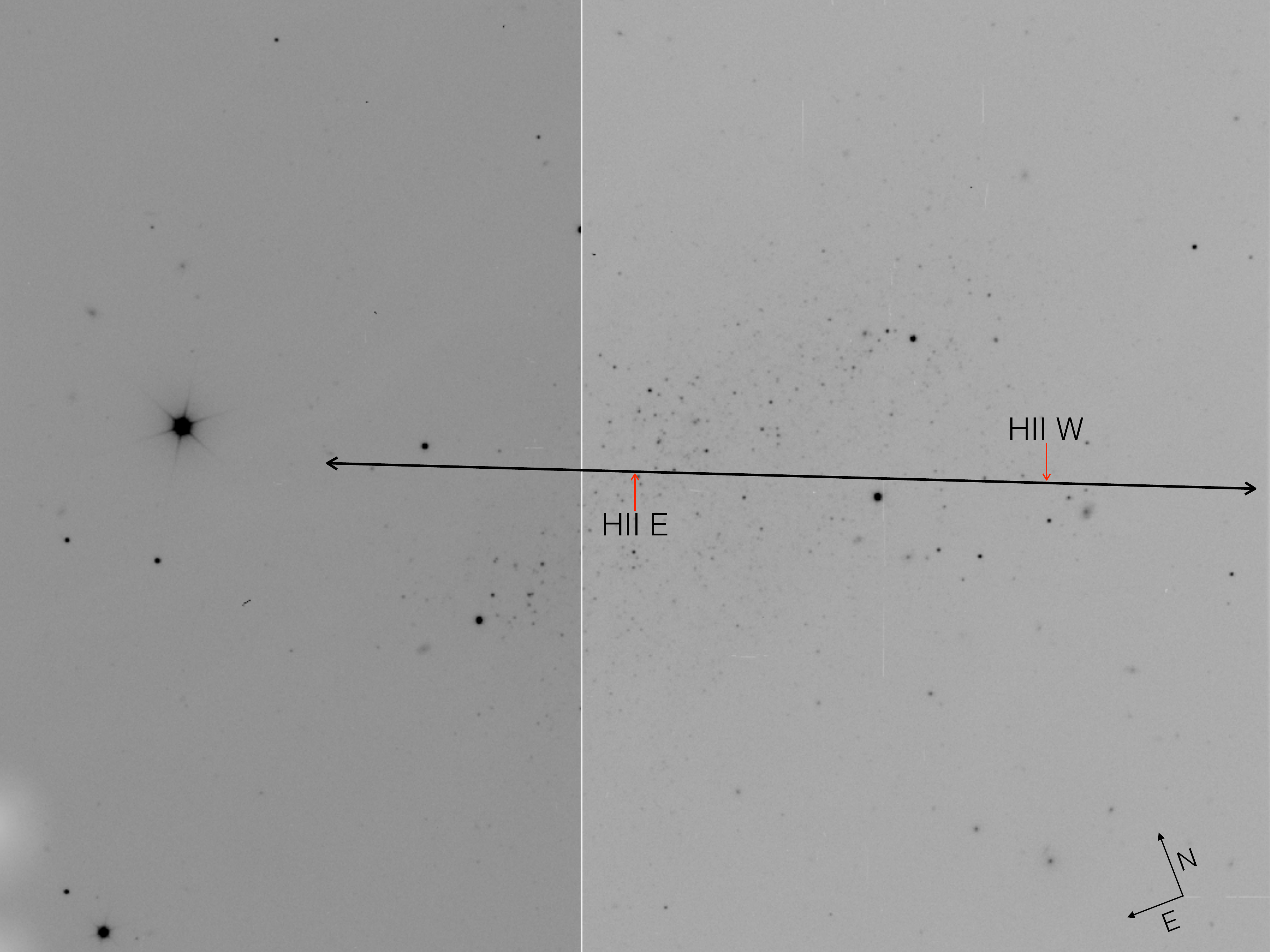}
\caption{Image of Leo A, obtained with OSIRIS-GTC, showing the orientation of the slit used. Two \ion{H}{II} regions (West and East) were included in the slit.\label{fig:slit}}
\end{figure*}

We observed two \ion{H}{II} regions in Leo~A using the Optical System for Imaging and Low Resolution Integrated Spectroscopy \citep[OSIRIS;][]{CepaAguiar2000,CepaAguiar2003} spectrograph attached to the GTC, at the Observatorio del Roque de los Muchachos (ORM) in La Palma, Spain.
The observations were carried out in service mode using the long-slit mode of OSIRIS during the nights 20 and 25 of April, 2015 under the Spain-Mexico collaborative GTC program {\sc GTC5-15AIACMEX}, that was awarded with five hours of observing time. The observing time was divided in five observing blocks (OB), one hour each, of which 45 min were used for scientific observations.   The 2D frames obtained with OSIRIS have a Field-of-View  of 7.8$\times$7.8 arcmin (unvignetted), and  data are deposited on an array of 2 CCDs. It is recommended to locate the targets on  CCD2 (CCD at West, see Fig. \ref{fig:slit}) to avoid shadowing on the border of CCD1.  A binning of 2$\times$2 was applied during the observations.

The slit size was 7.4 arcmin long and 1.8 arcsec wide and it was oriented as shown in Fig.~\ref{fig:slit}. The grism R1000B was used in order to cover the wavelength range from 3630 to 7500 \AA. This grism has a spectral dispersion  D=2.12 \AA/pix and, for the slit of 1.8 arcsec, it provides an spectral resolution of R=612 at 5460 \AA.

Due to the faintness of the \ion{H}{II} regions, a blind offset from a nearby star was necessary to place the objects in the slit; in that way the slit was centred at the coordinates $\alpha$=09:59:21.30; $\delta$=+30:44:20.0 (J2000), covering simultaneously two \ion{H}{II} regions (designed \ion{H}{II} West and \ion{H}{II} East, see Fig.~\ref{fig:slit}). The five OBs were obtained with the slit oriented in such a way. The science exposure time for each OB was of 2700 s, therefore the total scientific exposure time was 3.75 hrs. 

Coordinates for the \ion{H}{II} regions are: \ion{H}{II} West $\alpha$=09:59:17.2; $\delta$=+30:44:07 (J2000) and \ion{H}{II} East $\alpha$=09:59:24.5; $\delta$=+30:44:59 (J2000). Due to the characteristics of the nebular spectrum of \ion{H}{II} West we can identify it as the \ion{H}{II} region (-101-052) reported by \citet{vanZeeSkillman2006}; on the other side the \ion{H}{II} East has coordinates that match within uncertainties with those of SHK 1 by \citet{StrobelHodge1991}.

The observations were obtained at airmasses between 1.12 and 1.47 (Table \ref{observation_log}). Taking into account these values of airmass and, in particular, the slit width of 1.8 arcsec, the effects of differential atmospheric extinction are low-enough to not affect the observations. This was verified by comparing the extracted spectra from the lowest and highest airmasses OBs (see Table \ref{observation_log}). In both cases the line ratio [\ion{O}{II}] $\lambda$3727/H$\alpha$ is similar.

\begin{figure*} 
\includegraphics[width=1.5\columnwidth]{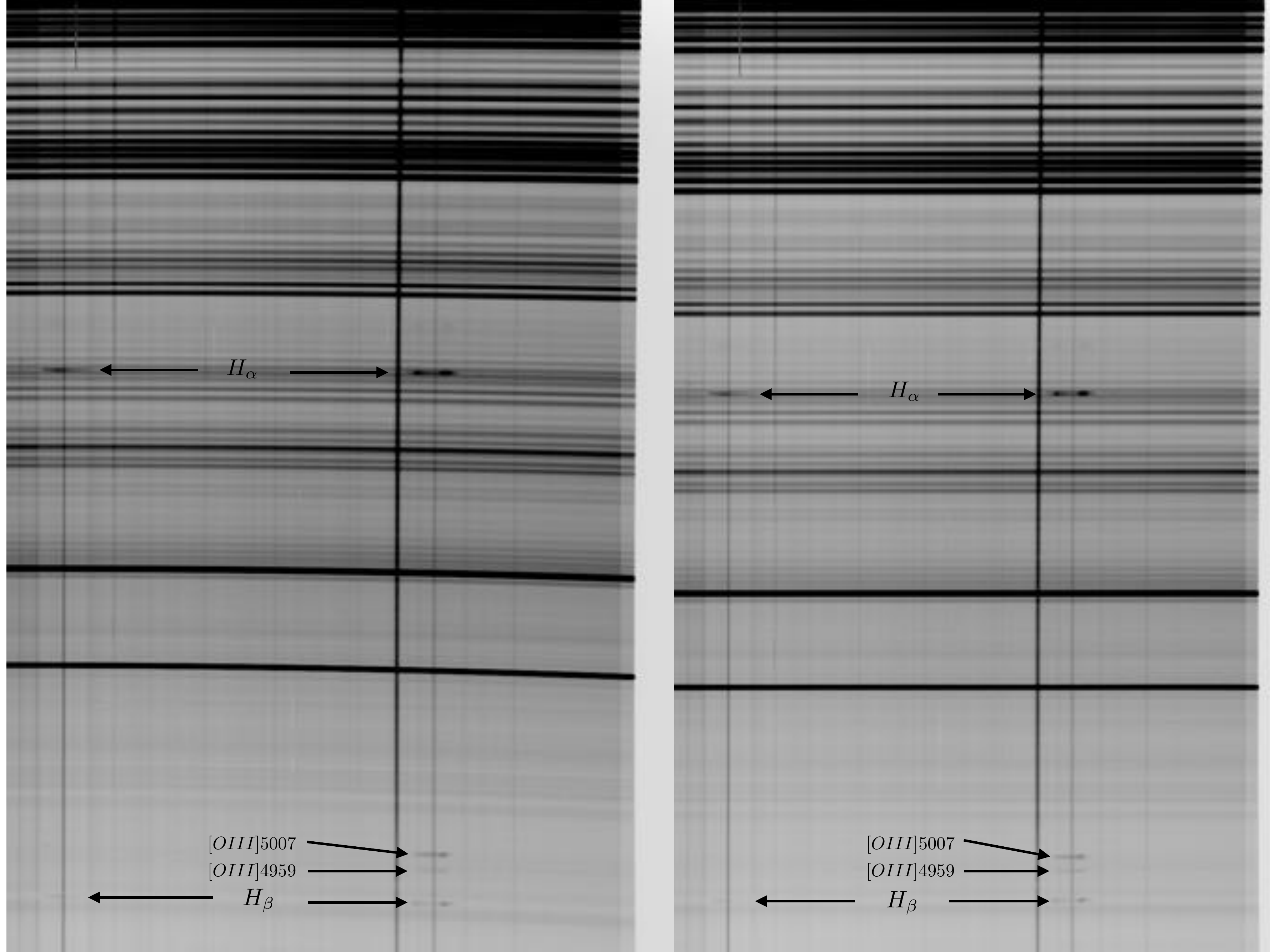}
\caption{ The observed (left) and rectified (right) 2D frames are shown.
\label{fig:rectified}}
\end{figure*}

The five 2D frames were aligned and combined in order to obtain the maximum signal-to-noise for the 2D spectra.  Data were reduced using IRAF\footnote{IRAF is distributed by the National Optical Observatories which is operated  by  the Association of Universities for Research in Astronomy, Inc.,  under cooperative agreement with the National Science Fundation.} packages and routines. Observed 2D frames were  bias subtracted and flat fielded  based on dome flats. A Hg-Ar lamp was used to calibrate the wavelength and to rectify the two-dimensional images. Fig. \ref{fig:rectified} left shows the 2D image with the spectra previous to the reduction and wavelength calibration and Fig. \ref{fig:rectified} right shows the 2D images after data reduction. 

\begin{table}
\begin{tabular}{ccccc} \hline
OB	& Obs. Date 		& Intial Obs. Time 	& Airmass \\ \hline \hline
OB1 	& 20/04/2015 	& 23:11:59 	& 1.12  \\
OB2 	& 21/04/2015 	& 00:00:54 	& 1.26  \\
OB3 	& 21/04/2015 	& 00:47:37 	& 1.47  \\
OB4 	& 25/04/2015 	& 23:02:42 	& 1.14  \\
OB5 	& 25/04/2015 	& 23:49:20 	& 1.29  \\ \hline
\end{tabular}
\caption{Log of observations. Position angle (PA) of the slit for all OBs is of 69 deg. Seeing was between 0.8 and 1 arcsec.}
\label{observation_log}
\end{table}

\begin{figure*} 
\includegraphics[width=0.8\columnwidth]{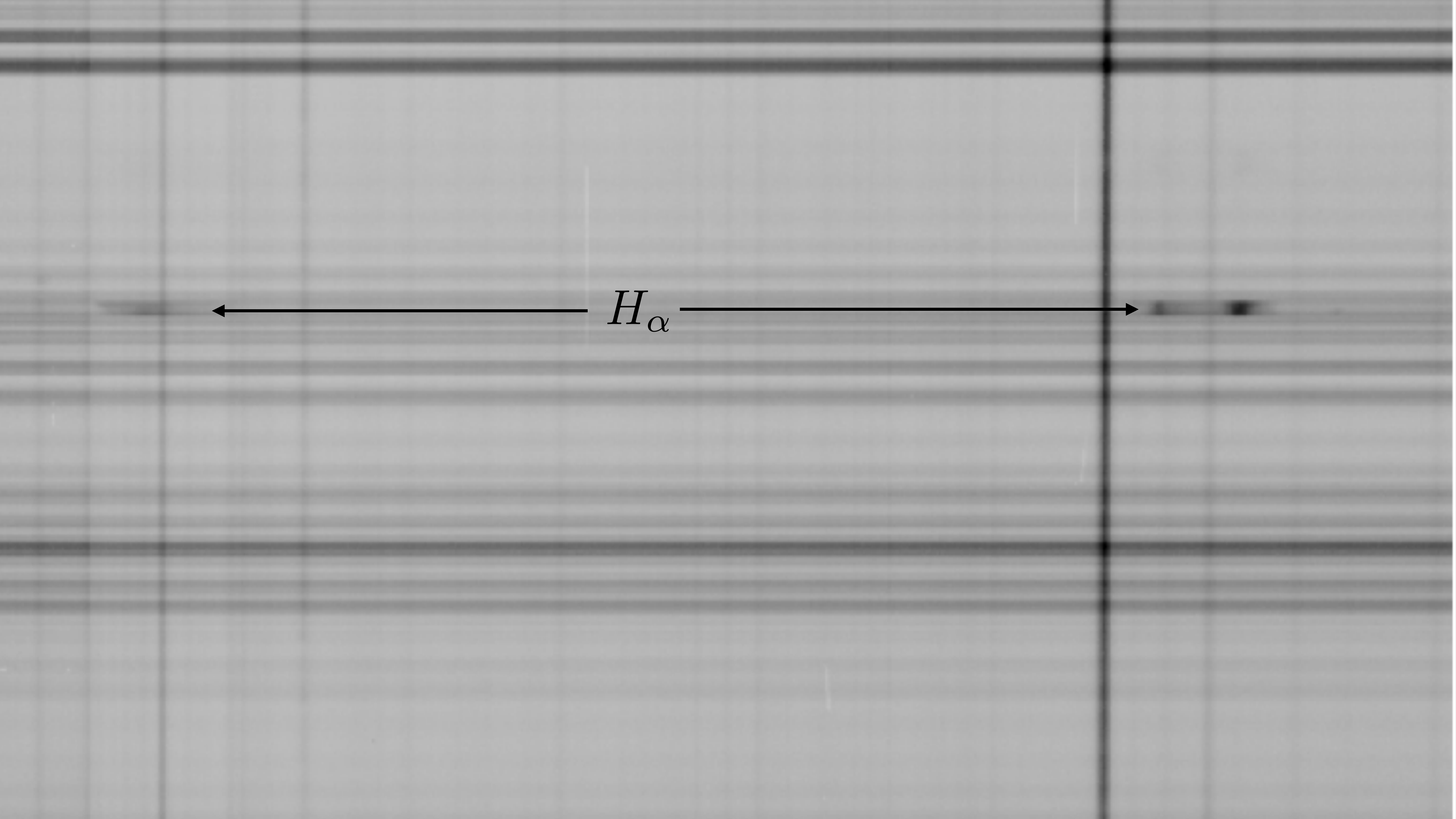}
\includegraphics[width=0.8\columnwidth]{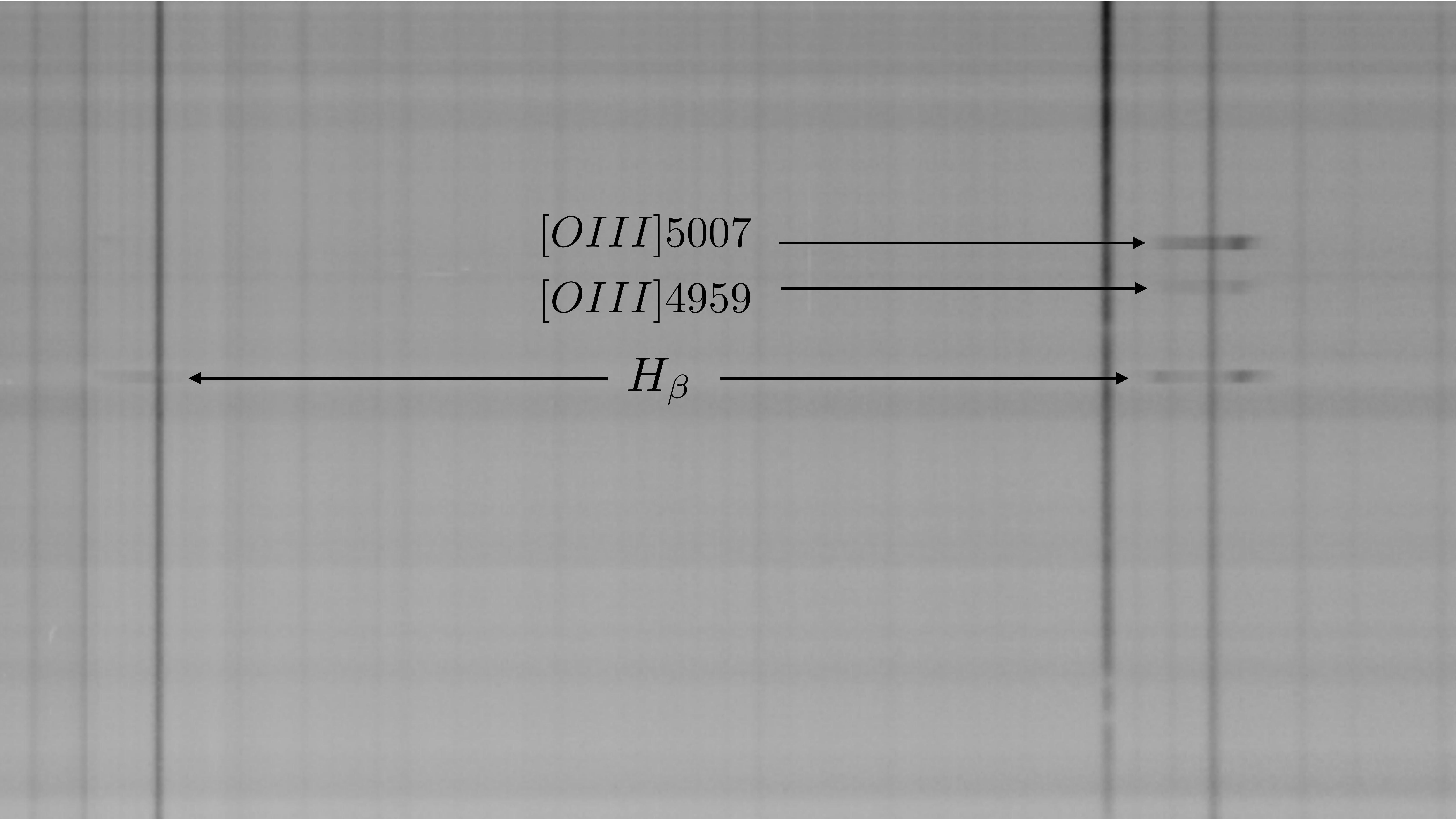}
\caption{Images with 2D calibrated spectra for the two \ion{H}{II} regions are shown. In the left the zones around H$\alpha$ are presented. In the right, the zones around H$\beta$ and [\ion{O}{III}] $\lambda$5007 are found. \ion{H}{II} region West is the one to the right in the images.
\label{fig:images}}
\end{figure*}

Fig. \ref{fig:images} left shows the zone around H$\alpha$ line while Fig. \ref{fig:images} right shows the zone around H$\beta$ for the two \ion{H}{II} regions.
In this figure it is observed  that both \ion{H}{II} regions are very extended. In both cases a faint stellar continuum is detected in the centre of the \ion{H}{II} emission, probably due to the ionising central stars.  In the right image it is appreciated that the [\ion{O}{III}]  $\lambda\lambda$5007 and 4959 lines only appear in the brightest region, \ion{H}{II} West.  In the H$\alpha$ images, the size of the \ion{H}{II} regions were measured to be approximately 2.8 and 1.5 pc (West and East respectively) by assuming a distance of 800 kpc to Leo A. 

One-dimensional spectra  were extracted for the  \ion{H}{II} regions from the rectified two-dimensional image and were corrected by effects of atmospheric extinction. Extraction was performed  trying to make a careful sky subtraction specially in the red zone where many sky lines appear. The sky was subtracted by fitting a low order polynomial. 

For the brightest region (\ion{H}{II} West) two extraction windows were used,
one at each side of the central star. This is a low excitation region, showing faint [\ion{O}{iii}] $\lambda\lambda$5007 and 4959 lines. Besides, the lines [\ion{O}{ii}] $\lambda$3727, [\ion{N}{ii}] $\lambda$6583 and [\ion{S}{ii}]$\lambda\lambda$6717,6731 are clearly seen. The zone east to the central star has better signal to noise, therefore this zone was used for the analysis of the emission lines. 

The second \ion{H}{II} region, \ion{H}{II} East, is much fainter and of lower excitation, only lines of the H Balmer series together with [\ion{O}{ii}] $\lambda$3727, [\ion{N}{ii}] $\lambda$6583 and [\ion{S}{ii}] $\lambda\lambda$6717,6731 lines are detected.

\begin{figure*}
\includegraphics[width=1.04\columnwidth]{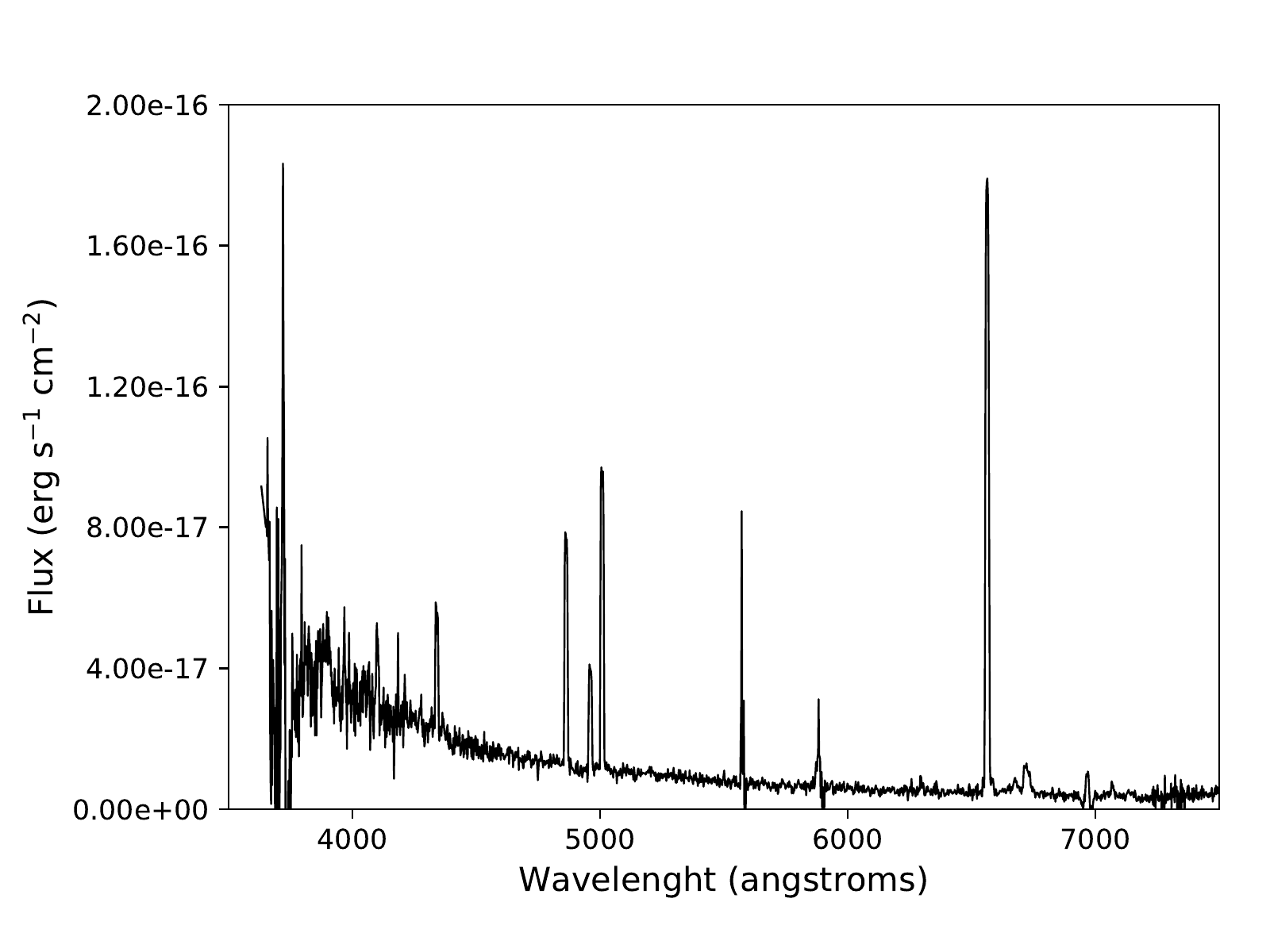}
\includegraphics[width=1.04\columnwidth]{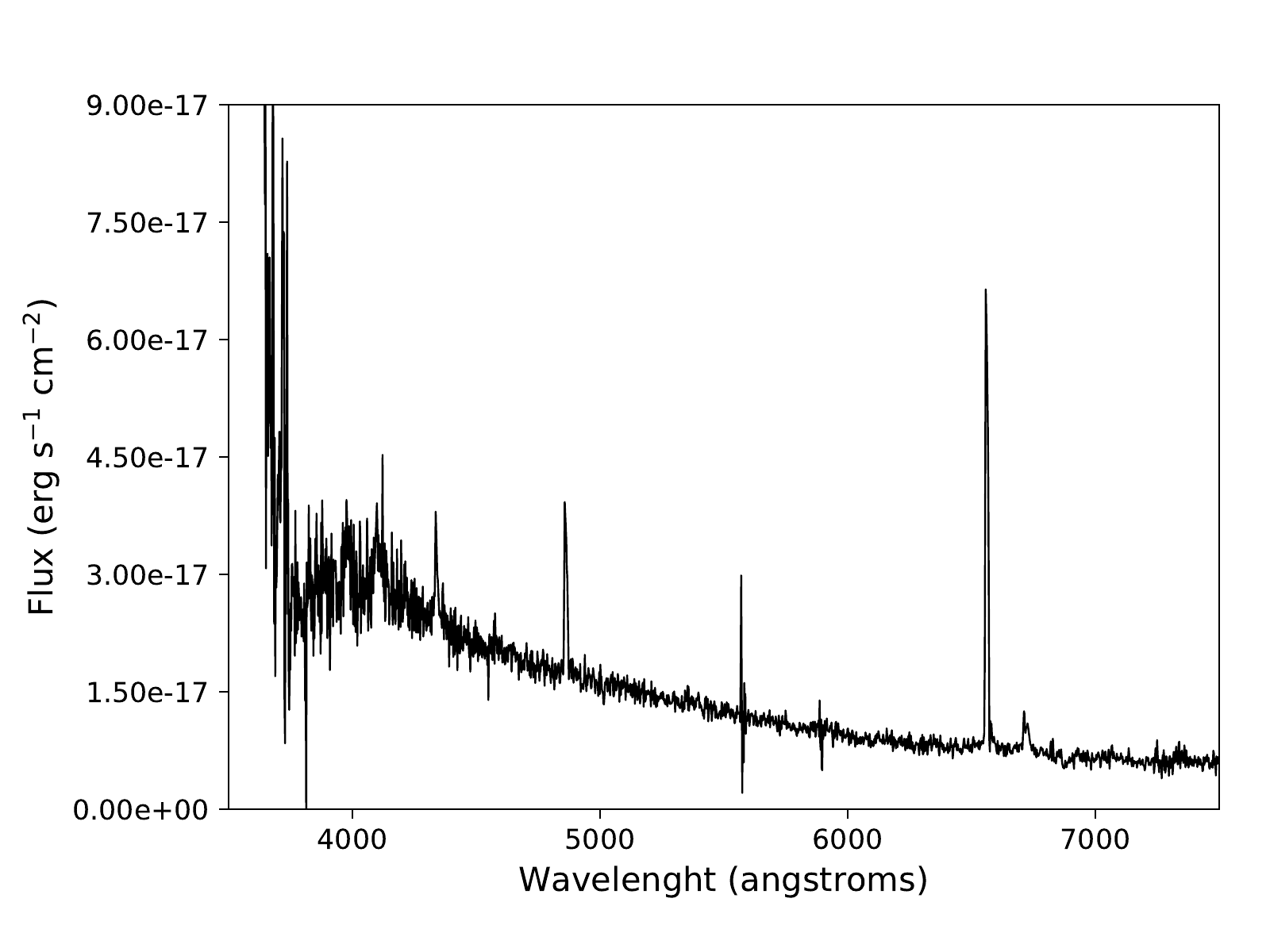}
\caption{Calibrated spectra of both \ion{H}{II} regions. The brightest one, \ion{H}{ii} West (left) presents [\ion{O}{III}]  $\lambda\lambda$5007 and 4959 lines, besides [\ion{O}{II}]  $\lambda$3727. \ion{H}{II} region East  (right) only shows the H Balmer lines, [\ion{O}{II}] $\lambda$3727 and very faint [\ion{N}{II}] and [\ion{S}{II}]  lines. 
\label{fig:spectra}}
\end{figure*}

 The standard star  Hiltner 600 was used for flux calibration. Extracted calibrated spectra are shown in Fig. \ref{fig:spectra}.

\section{Line intensities,  physical conditions and ionic abundances in H\,{\sc ii} regions}

All available line fluxes were measured by integrating all the emission in the line and fitting the continuum by eye. Line flux errors were calculated by using the \citet{TresseMaddox1999} expression shown in Eq. (\ref{eq:treese}):
\begin{equation}
\sigma_{I} = \sigma_{c}D\sqrt{2N_{pix}+\frac{EW}{D}},
\label{eq:treese}
\end{equation}
where D is the spectral dispersion in \AA/pix, $\sigma_c$ is the mean standard deviation per pixel of the continuum on each side of the line, and $N_{pix}$ is the number of pixels covered by the line.

Line fluxes relative to H$\beta$  are listed in Table \ref{table:intensities},  column 4. In column 5 we present the line uncertainties derived by propagating the line flux errors of the line and of H$\beta$. 
Notice that in this table, the \ion{He}{i} line $\lambda$5876 is not listed because, although it is visible in the spectrum of \ion{H}{II} West, it is blended with the atmospheric Meinel rotation-vibration bands of OH: 8$-$2 Q2 (0.5) at 5887.129 and 8$-$2 Q1 (0.5) at  5888.139 \AA\ \citep{OsterbrockFulbright1996}; whereas for \ion{H}{II} East the line is not detectable.

Spectral fluxes  were dereddened by using the parametrisation of the Galactic standard extinction law by \citet{Seaton1979} and the logarithmic reddening corrections, c(H$\beta$), which were derived in each case from the H$\alpha$/H$\beta$ Balmer ratio by assuming case B recombination line intensities \citep{StoreyHummer1995}. 
 c(H$\beta$) and the observed H$\beta$ flux values are included at the end of  Table \ref{table:intensities}. Errors in c(H$\beta$) values were derived by propagating the line flux errors of H$\alpha$ and H$\beta$ through the c(H$\beta$) expression.

In column 6  of Table \ref{table:intensities}, the dereddened line intensities are listed. 
Uncertainties of dereddened line intensities are presented in column 7, these were calculated by propagating the errors of c(H$\beta$) and the line flux errors through the dereddening correction.

\subsection{Physical conditions and ionic abundances}

Physical conditions (density and temperature) and ionic abundance determinations were performed by using the code  PyNeb v.1.1.1 \citep{LuridianaMorisset2015} and the atomic parameters listed in Table \ref{table:atomic-data}.

The density sensitive [\ion{S}{II}]  $\lambda\lambda $6717/6731 line ratio was detected in both \ion{H}{II} regions and it was used to determine electron densities. Both regions have low-density with n$_e$ of about 100 cm$^{-3}$. The faint auroral line [\ion{O}{III}]  $\lambda$4363 is barely detected in \ion{H}{ii} region West, thus the [\ion{O}{III}]  $\lambda\lambda$4363/5007 line ratio was used to calculate the electron temperature in this nebula, although the uncertainty is large.  

The derived  physical conditions are presented in Table \ref{table:abundances}.  The measured temperature for region \ion{H}{II} West  is about 17000$\pm$3000 K. As it was commented above, the large uncertainty is owing to the faintness of the auroral [\ion{O}{III}]  $\lambda$4363 line.

These physical conditions were employed to calculate ionic abundances of the available ions. A two-temperature zone model was adopted. For the O$^{+2}$ zone the [\ion{O}{III}]  temperature was used. For the lower ionisation species (O$^+$, N$^+$ and S$^+$) a temperature computed from the expression by \citet{CampbellTerlevich1986},  presented in Eq. \ref{eq:Campbell}, was used.

\begin{equation}  
  T_{e}[\ion{O}{II}]  = 0.7~T_{e}[\ion{O}{III}]  + 3000 K 
  \label{eq:Campbell}
\end{equation}

This relation, derived by \citet{CampbellTerlevich1986} from \citet{Stasinska1982} photoionisation models of \ion{H}{II} regions, allows the construction of a two-temperature zone model in objects in which only one of these temperatures can be determined from the observational data.

The results for ionic abundances are listed in Table \ref{table:abundances}, and were used to determine total abundances as it is explained in \S 4.

The \ion{He}{I} $\lambda$6678 line, detected in \ion{H}{ii} West, was used to determine the He$^+$ ionic abundance because it was not possible to measure the \ion{He}{I} $\lambda$5875 line flux. It is important to mention that the detection of the He$^+$ lines is indicative of the stellar effective temperature. When these lines are visible the temperature is larger than 35,000 K  \citep{SkillmanBomans1997}.


\begin{table*}
\centering
\begin{tabular}{lcc} 
\hline
Ion 			& Transition Probabilities 								& Collisional Strengths \\ \hline \hline
Ar$^{+2}$ & \citet{Mendoza1983, KaufmanSugar1986} 	& \citet{GalavisMendoza1995} \\
N$^+$ 		& \citet{FroeseFischerTachiev2004} 				& \citet{KisieliusStorey2009} \\
Ne$^{+2}$ & \citet{GalavisMendoza1997} 						& \citet{McLaughlinBell2000} \\
O$^{+}$ 	& \citet{FroeseFischerTachiev2004} 				& \citet{KisieliusStorey2009} \\
O$^{+2}$ 	& \citet{FroeseFischerTachiev2004} 				& \citet{StoreySochi2014} \\
S$^{+}$ 	& \citet{PodobedovaKelleher2009}					& \citet{TayalZatsarinny2010} \\ \hline
\end{tabular} 
\caption{Atomic data used to calculate physical conditions and ionic abundances}
\label{table:atomic-data}
\end{table*}

\subsection{The PN}

To analise this object, we adopted the dereddened line intensities presented by  \citet{vanZeeSkillman2006} to calculate the physical conditions and chemistry of the PN. For consistency in these calculations we used the same procedure and atomic parameters used for our data for the \ion{H}{ii} regions. 

In Table \ref{table:intensities} the line intensities and the associated uncertainties are listed. It is important to note that the PN does not show the low ionisation line intensities. No [\ion{O}{ii}] and [\ion{S}{ii}] lines were detected and the [\ion{N}{ii}] $\lambda$6583 is  barely visible, which indicates that this PN is a density-bounded nebula. The lines from twice ionised species however are intense.

The auroral line [\ion{O}{ii}] $\lambda$4363 is clearly detected, therefore the [\ion{O}{iii}] $\lambda\lambda$4363/5007 line ratio can be  safely used to determine the electron temperature. A value  of about 20900$\pm$930 K was derived from this line ratio.  On the other hand no density-sensitive line ratio was detected then we adopted a reasonable value n$_e$ $\sim$ 1000 cm$^{-3}$ for the density in this object.

Ionic abundances were determined  by assuming a two-temperature zone model, the same as in \ion{H}{ii} regions. The results are shown in Table \ref{table:abundances}.

This nebula shows a relatively intense \ion{He}{ii} $\lambda$4686 line, therefore it is a highly ionised nebula with a hot central star. The existence of the ion He$^{+2}$ in the nebula indicates the presence of O$^{+3}$ and other highly ionised species. This should be taken into account in the calculus of the total abundances.

\section{Total abundances for the H\,{\sc II} regions and the PN}

To calculate total abundances from ionic abundances, the unobserved ions should be included. For this, it is common to use the so called Ionisation Correction Factors (ICFs). When analising the chemistry in the Orion Nebula, \citet{PeimbertCostero1969} proposed several ICFs based on considering similarity in ionisation potentials by assuming that ions of the same ionisation potential coexist in the same zone of the ionisation structure. Many authors have revised the ICFs proposed by these authors recomending slightly different ICFs \citep[see e.g.,][K\&B94]{KingsburghBarlow1994}. 
Recently, \citet[D-I2014]{Delgado-IngladaMorisset2014} have calculated detailed ICFs for elements in planetary nebulae, based on the analysis of a large number of ionisation structure models of different characteristics. In this work we have decided to use the ICFs for \ion{H}{ii} regions and  PNe, currently used in the literature which are mentioned in the following.

 The He abundance deserves a special explanation. Inside the photoionised regions the presence of neutral He could be expected in low ionisation nebula. The correction for such an ion is not, so far,  well determined, therefore for calculating the total He abundance we only consider the presence of He$^+$ and He$^{+2}$. This is a good approximation for the PN where no neutral He is expected due to its high ionisation degree, but for the \ion{H}{ii} regions this is only an upper limit of the total He abundance.

Total abundances of He, O, N and S in the \ion{H}{ii} region West and He, O, N, Ne, Ar, and S  in the PN were derived by using the ICFs indicated in the following expressions:

\defcitealias{Delgado-IngladaMorisset2014}{D-I2014} 
\defcitealias{KingsburghBarlow1994}{K\&B1994} 

\smallskip

For the \ion{H}{II} regions:

\noindent - He/H = He$^+$/H$^+$.

\noindent - O/H = (O$^+$ + O$^{+2}$)/H$^+$.

\noindent - N/O = N$^+$/O$^+$, \citetalias{KingsburghBarlow1994}. 

\noindent - S/H = 2.5 S$^+ \times$ (O/O$^+)^2$,  \citet{Stasinska1978}. 

\smallskip

For the PN:

\noindent - He/H = He$^+$/H$^+$ + He$^{+2}$/H$^+$

\noindent - O/H = (O$^+$ + O$^{+2}$)/H$^+ \times 10^{ (0.068 \nu ^2 + 0.06 \nu) / (0.34 - 0.08\nu) }$, \noindent where $\nu$=He$^{+2}$/(He$^{+}$+ He$^{+2}$),  \citetalias{Delgado-IngladaMorisset2014} \footnote{This is the most recent ICF available in the literature to correct the total O abundance for the presence of the unobserved ion O$^{+3}$. Previous ICFs by \citet{Torres-PeimbertPeimbert1977} and \citet{KingsburghBarlow1994} provide slightly different values for the correction. The results are similar, within uncertainties.}. 

\noindent - Ne/O = Ne$^{+2}$/O$^{+2}\times(w+(0.14/\nu+2 \nu^{2.7})^3 \times (0.7+0.2w-0.8w^2$)), where $w$=O$^{+2}$/(O$^+$ + O$^{+2})$,  \citetalias{Delgado-IngladaMorisset2014}. 


\noindent - Ar/H = 1.87 Ar$^ {+2}$ /H$^+$, \citetalias{KingsburghBarlow1994}.

\smallskip

\noindent The results are presented in Table \ref{table:abundances} for the three nebulae.
\smallskip

Due to the large uncertainty in temperature measured for \ion{H}{II} region West, which produces a large uncertainty in the derived abundances we also used the strong-line method called ONS, proposed by \citet{PilyuginVilchez2010} to determine the total abundances of O/H and N/H for this region. 

The ONS method relates the measured intensities of [\ion{O}{III}] $\lambda\lambda$4959,5007, [\ion{O}{II}] $\lambda$3727, [\ion{N}{II}] $\lambda$6548+83, [\ion{S}{II}] $\lambda$6717+31 and H$\beta$ lines with the total abundances of O/H and N/H. The calibration of this method is based on a sample of 118 \ion{H}{II} regions from the Galaxy disc, discs of nearby spiral galaxies and irregular galaxies, with accurate measured electronic temperatures and spanning a large range of metallicities. To determine the O/H and N/H abundance, ONS method assigns different relations for cold, hot and warm \ion{H}{II} regions, that are distinguished by [\ion{N}{II}] $\lambda\lambda$6548+83/H$\beta$ and [\ion{S}{II}] $\lambda\lambda$6717+31/H$\beta$ lines ratios. Under \citet{PilyuginVilchez2010} definition, \ion{H}{II} region West is a hot region, so their abundances are determined following the respective abundance-intensities relation. 
\citet{PilyuginVilchez2010} claim that ONS method gives realistic values of \ion{H}{II} region abundances because it shows very good agreement, better than 0.075 dex for O and 0.05 dex for the N abundances, with abundances derived from the direct method based on measured electron temperatures. This idea is supported by \citet{Arrellano-CordovaRodriguez2016}, who recommend the use of the ONS method when no temperature determinations are possible or when the available determinations are of poor quality.

The results of the ONS method for \ion{H}{II} West are  included in Table \ref{table:abundances}. For this region, direct method and strong-line method results agree within 0.01 dex for O and within 0.16 dex for N. The ONS method could not be applied to the \ion{H}{II} region East, due to the absence of a reliable measurement of the [\ion{O}{III}] $\lambda$5007 line. 

In order to double-check the abundances derived from direct- and ONS-methods, we computed photoionisation models trying to derive robust chemical abundances for both \ion{H}{II} regions and the planetary nebula.

\section{Photoionisation structure models}

The code CLOUDY v.17 \citep{FerlandChatzikos2017}, available from the literature, was used to compute photoionisation models for both \ion{H}{II} regions and the planetary nebula. The pyCLOUDY library \citep{Morisset2013} was used to construct the input files and to analyse the models results.
With these models we intend to reproduce the ionisation structure, the physical conditions and the line intensities of the nebulae.

CLOUDY is a code  of spectral synthesis which calculates the physical conditions in interstellar clouds which can be exposed to an external radiation field. Given some initial parameters such as the intensity and spectral distribution of the radiation field, the nebular geometry, and the chemical composition of the gas (the first thirty lightest elements can be included),  the code computes the ionisation equilibrium, the ionisation degree, the statistical equilibrium, the energy and charge conservation, the physical conditions (electron density and temperature) and the full spectrum including thousands of lines,  by assuming stationary equilibrium. For this calculation the code considers  all the ionisation states and recombination mechanisms (charge exchange, radiative and dielectronic recombination, phototoionisation and collisional excitation). 

To calculate models for our \ion{H}{ii} regions and the PN, the required input parameters are the stellar effective temperature T$_{eff}$, the stellar surface gravity $g$, and the ionising photons rate $Q(H^0)$ of the central star, as well as the nebular density structure and the nebular chemical composition. A grid of models was computed for each case varying these parameters. Because the real nebular structure is not known, we computed single models for the three objects considering  
 a single Str\"omgren sphere of constant density. The models are calculated from inside out, and finish when the electron temperature has decreased down to 4000 K  for the \ion{H}{II} regions, and 16500 K, for the planetary nebula (see below). For the three objects, the ionising Stellar Emission Distribution (SED) is simulated by using the stellar atmospheres grids available in CLOUDY for stars of metallicity 1/10 Z$_{\odot}$ (this is considering the low metallicity of this galaxy). For both \ion{H}{II} regions, the SED is simulated using the TLUSTY grids of O and B stars \citep[respectively]{LanzHubeny2003,LanzHubeny2007}, while for the planetary nebula, a stellar atmosphere grid of H-Ni metal-line blanketing  from \citet{Rauch2003} is used.

\subsection{Models for the H{\sc{II}} regions}
For the brigther and more ionised nebula, \ion{H}{II} West, we fit first the nebular ionisation degree given by the O$^+$/O$^{+2}$ ionic ratio. This helps to determine the stellar parameters. Afterwards we tried to fit the intensity of emission lines, by adjusting the chemical abundances. For this region a star of  T$_{eff} \sim$ 35,600  K (spectral type O8V - O9V) was necessary to reproduce the nebular ionisation structure. 

	For the lower ionised region, \ion{H}{II} East, which shows only an upper limit to the [\ion{O}{III}] $\lambda$5007 emission, a much colder star of T$_{eff} \sim$ 28,000 K (spectral type B0V - B1V) was required. The chosen  stars generate photoionised regions of radius similar to the ones observed (about 2.7 pc for \ion{H}{ii} West and 1.5 pc for \ion{H}{ii} East), when an electron density of 100 cm$^{-3}$ is assumed. 

The chemical abundances used initially for the model of \ion{H}{II} region West were those derived from observations. This only provides values for O/H, N/H and S/H. For the other elements (C, Ne, Ar and Si), we adopted the values of abundance ratios relative to O determined for the Small Magellanic Cloud (SMC)  by \citet{Garnett1999}. The SMC is a very well studied low-metallicity irregular galaxy in the Local Group and its chemical behaviour could be comparable to the one in the low-metallicity galaxy Leo A. In the adjustment of the model, fine-tuning of the abundances was necessary.

For the model of \ion{H}{II} East, due to the absence of observational constraints, the chemical abundances were those used for the \ion{H}{II} West model except for O and S, which needed to be adjusted to fit the observed line intensities.

The results for the best models as well as the stellar parameters used, are listed in Table \ref{table:abundances}.

\subsection{Models for the PN}
In the case of the planetary nebula, as we say above, the emission of the low ionisation species S$^+$ and O$^+$ are not detected in the nebular spectrum presented by \citet{vanZeeSkillman2006}.
Only a faint  emission of [\ion{N}{ii}]  $\lambda$6584 is present, therefore the nebula seems to be density-bounded. On the other hand, the \ion{He}{II} $\lambda$4686 line is intense in this nebula, therefore this is a highly ionised nebula. To construct a photoionisation model, initially we adopted a high-temperature star (T$_{eff}$ larger than 100,000 K and log(g)=7.0), an electron density of 1000 cm$^{-3}$, and used the chemical abundances as derived from \ion{H}{ii} regions. 

To fit the density-bounded restriction, the photoionisation code was stopped when the nebular temperature declined to less than 16500 K, which would be the temperature of the low ionisation zone estimated with \citeauthor{CampbellTerlevich1986} relation (Eq. \ref{eq:Campbell}), from the observed T(\ion{O}{iii}) temperature.   Once this restriction was imposed, the stellar parameters were determined by fitting the nebular ionisation degree given by the He$^{+}$/He$^{+2}$ ionic ratio. The parameters that best fit the ionisation degree are T$_{eff}$= 125,000 K, log(g) = 7.0 and log of the ionising photon rate equal to  48. Similarly to the models for the \ion{H}{II} regions,  the emission line intensities were fitted by adjusting the gas chemical composition. The results predicted by the best model are listed in Table \ref{table:abundances}.

\section{Discussion and Conclusions}

In this work we have derived the chemical composition in two \ion{H}{II} regions in Leo A, with different methods: the direct method (by using a very uncertain electron temperature measured in one of the \ion{H}{II} regions), the ONS method of strong lines \citep{PilyuginVilchez2010} and through the computation of photoionisation models. For the highest excitation \ion{H}{II} region (\ion{H}{ii} West) all the methods provide similar chemical abundances for O and N with values 12+log(O/H) $\sim$ 7.4 - 7.5 and  log(N/O) $\sim -$1.7. The fainter \ion{H}{ii} region East does not show [\ion{O}{iii}] lines, therefore no direct method or ONS method can be applied, and its chemical composition was derived  from a photoionisation model,  giving 12+log(O/H) $\sim$ 7.4 and log(N/O) $\sim$ $-$1.6.
 
 Thus, it is confirmed that Leo A is one of the most metal poor  galaxies in the Local Group, where its present ISM has 12+log(O/H) $\sim$  7.4$\pm$0.2. Conclusively this galaxy has evolved very slowly  in agreement with the star formation history derived by \citet{ColeSkillman2007} which includes initial 5 Gyr of quiescence. The  log(N/O) = -1.6 is extremely low compared with the values for evolved galaxies as the Milky Way which presents log(N/O) of about -0.8 \citep{EstebanGarcia-Rojas2018}, but it is consistent with values found in irregular galaxies. From a sample of  more than 30 low-metallicity irregular galaxies, \citet{Garnett1990} found that the ratio log(N/O) varies from $-$1.7 to $-$1.4, with an average log $<$N/O$>$ = $-$1.47$\pm$0.10, therefore the low N/O ratio in Leo A is similar to the values for low-metallicity irregular galaxies.
 
The photoionisation models for the two analised \ion{H}{II} regions were computed with the code CLOUDY and the best fit models  show that the ionising stars in these \ion{H}{II} regions correspond to a star with T$_{eff}$ of about 35600 K, and ionising photon rate of 10$^{48.9}$ photons s$^{-1}$ for the more ionised \ion{H}{ii} region West, and to a star of T$_{eff}$=28000 K and ionising photons rate of 10$^{48.0}$ photons s$^{-1}$ for the less ionised \ion{H}{ii} region East. 

It is interesting to notice that \ion{H}{ii} West is the highest excitation \ion{H}{ii} region in Leo A, considering the 4 \ion{H}{ii} regions studied by \citeauthor{vanZeeSkillman2006}, and its central star has a mass of about 25 M$_\odot$. This indicates that a low-mass galaxy like Leo A, with a present SFR of about 10$^{-4}$ M$_\odot$ yr$^{-1}$  \citep{ColeSkillman2007} is capable of forming high mass stars, similarly to what occurs in the very low-mass galaxy Leo P \citep{McQuinnSkillman2015}. However and although we have not analysed the whole sample of \ion{H}{ii} regions in the galaxy, in the present starburst in Leo A it appears that the most massive star formed has mass greater or equal 25 M$_\odot$ and this is in agreement with the chemical evolution model by \citetalias{Hernandez-MartinezCarigi2018} where the upper limit required for  the IMF for Leo A is 40 M$_\odot$.

\medskip 
The only PN known in this galaxy is also a very metal poor object. Its O/H abundance as derived from the direct method and the photoionisation model  is 12+log(O/H) $\sim$ 7.34 -- 7.45. 
The best model required a central star with T$_{eff}$ of 125,000 K, log(g) = 7.0 and ionising photon rate of 10$^{48}$    photons s$^{-1}$.

The PN O/H abundance ratio is similar to the abundances of the \ion{H}{ii} regions and the same is true for other elements like Ne, Ar, and S. However the N/O abundance ratio is larger than in \ion{H}{ii} regions, with a value log(N/O) of $-$0.45. It is common to find that planetary nebulae have enriched their N abundances (also He and C)  due to dredge-up processes or Hot Bottom Burning. In this case the enrichment is similar to the N enrichment of a  disc PN in our galaxy \citep[see e.g.,][]{KingsburghBarlow1994}. 
The N/O value is not sufficiently high to declare this PN as a Peimbert Type I PN which are PNe with N/O larger than 0.5 and also seem  to be He-enriched \citep{Peimbert1978}. The chemical abundances for this PN can be compared to the values predicted by stellar evolution models for low-intermediate mass stars by, e.g., \citet{Karakas2010} and \citet{FishlockKarakas2014} and it is found that the central star should have had an initial mass lower than about 2.0 M$_{\odot}$ in order to produce the observed chemical abundances in the PN. 

\section{acknowledgements}

This work is based on observations made with the Gran Telescopio Canarias (GTC), installed in the Spanish Observatorio del Roque de los Muchachos of the Instituto de Astrof{\'\i}sica de Canarias, in the island of La Palma, Spain. 
This work received partial support from the DGAPA-UNAM, Mexico under grant PAPIIT IN103117.  F.R.E. acknowledges scholarship from CONACyT, Mexico. J.G.R. acknowledges support from a Severo Ochoa excellence program (SEV-2015-0548) and support from the Spanish Ministerio de Econom\'ia y Competividad under project AYA2015-65205-P.




\bibliographystyle{mnras}
\bibliography{ref_leo_a} 




\begin{landscape}
\begin{table}
\caption{Line intensities }
\begin{tabular}{lrrrrrrrrrrrrrrrr}
\hline\hline
       & & &      \multicolumn{5}{c}{ \ion{H}{II}  West}  &  \multicolumn{5}{c}{    \ion{H}{II} East }   &  \multicolumn{3}{c}{PN}\\
\cmidrule(lr){4-8} \cmidrule(lr){9-13} \cmidrule(lr){14-16} 
Ion     &             $\lambda_0$    &    f($\lambda$) &     F/H$\beta$ &     $\Delta$  & I/H$\beta$    & $\Delta$ &       Model &         F/H$\beta$ &   $\Delta$ &   I/H$\beta$   &  $\Delta$& Model  &  I/H$\beta^a$    &$\Delta$ & Model\\
\hline
[\ion{O}{II}]    &               3727  &    0.26  &    1.64   &   0.45  &    1.75   &   0.48    &       2.05    &  1.50   &   0.45  &    1.58   &   0.47   &     1.61 & $<$0.08 & --- & 0.08 \\

[\ion{Ne}{iii}]   	&	3869 	&	---		& 	--- & --- & --- & ---&--- & --- & --- & --- & --- & --- & 0.27 & 0.02 & 0.26\\
H$\delta$      		& 	4102  	&	0.17  &    0.25   &   0.06  &    0.26   &   0.06    &       0.26    &  0.22   &   0.06  &    0.23   &   0.06   &     0.26 & --- & --- & ---\\
\ion{C}{II}, \ion{O}{II}				& ---  	   	& 	0.25  &    0.06   &   0.02  &    0.07   &   0.02 &--- &--- &--- &---&--- &--- &--- &--- &--- \\
H$\gamma$   	& 4340 		&	0.35  &    0.45   &   0.03  &    0.49   &   0.03    &       0.47 & 0.41 &0.05 & 0.42 & 0.05 & 0.47 & 0.44 &0.02 & 0.48\\
$[\ion{O}{III}] $   & 4363 		& 	0.07 	& 0.03 & 0.01 & 0.03 & 0.01 & 0.02 & --- & --- & --- & --- & --- & 0.14 & 0.01 & 0.12\\
\ion{He}{i}   		&4471 		& ---  	&--- &--- & --- & --- & --- & --- & --- &--- & --- & --- & 0.05 & 0.01 & 0.04\\
\ion{He}{ii}  		& 4686 		& --- 		& --- & --- & --- & --- & --- & --- & --- & --- & --- & --- & 0.21 & 0.02 & 0.19  \\
H$\beta$      		&	4861  	& 	0.00  &    1.00   &   0.03  &    1.00   &   0.03    &       1.00    &  1.00   &   0.05  &    1.00   &   0.05   &     1.00  & 1.00 &0.04 & 1.00\\
$[\ion{O}{III}] $   & 	4959  	&	-0.22	&    0.44   &   0.02  &    0.42   &   0.02    &       0.42  &   ---   &    --- &   ---   &  ---    & ---& 1.28 &     0.04   & 1.27\\
$[\ion{O}{III}]  $  &	5007		&	-0.33	&    1.35   &   0.03  &    1.24   &   0.05    &       1.24   &<0.06    &  ---   & <0.06    &  ---    &     0.00  & 3.78 & 0.12 & 3.78\\
$[\ion{O}{I}]  $    & 	6300  	& -0.28	&    0.02   &   0.01  &    0.01   &   0.01    &       0.00  &   ---     &   ---  &   ---   &  ---    & --- & ---  &--- 	&---   \\
H$\alpha$        	&	6563  	&	-0.32&    3.03   &   0.08  &    2.79   &   0.11    &       2.80  &   2.79   &   0.11  &    2.80   &   0.15   &     2.82  & 2.70 &0.13 & 2.75\\
$[\ion{N}{II}] $    & 	6583  	& -0.33 &    0.04   &   0.01  &    0.04   &   0.01    &       0.03   &   0.06   &   0.01  &    0.05   &   0.01   &     0.05 & 0.023 & 0.003 & 0.02\\
\ion{He}{I}       	& 	6678  	&-0.34 	&    0.05   &   0.01  &    0.04   &   0.01    &       0.03  &  ---    &   ---   &   ---   &   ---    &   0.00  & 0.04 &   0.004 & 0.04  \\
$[\ion{S}{II}] $  	& 	6717  	&  -0.34	&   0.13    &  0.01   &   0.12    &  0.01     &      0.09    &  0.11        &  0.01   &   0.11    &  0.01    &    0.09 & --- & --- &---\\
$[\ion{S}{II}] $  	&	6731  	&-0.34	&    0.10   &   0.02  &    0.09   &   0.02    &       0.07   &   0.07      &   0.01  &    0.07   &   0.01   &    0.07 & --- & ---	&--- \\
\ion{He}{I}       	&	7065  	&	-0.38	&    0.04   &   0.01  &    0.04   &   0.01    &       0.02  &   ---      &  ---   & ---     &  ---   &          & 0.13 & 0.01 & 0.05 \\

[\ion{Ar}{III}] &   7135   & --- 	& --- 	& ---	& --- 	&---	 &---	 &---	 &---	 &---	 &---	 &---	 & 0.01 &  0.003  & 0.01\\ \hline
c(H$\beta$) & &&                             0.11 &0.04                                      &&&&         0.08&0.05  & & & &   0.16 & 0.05\\
\multicolumn{3}{l}F(H$\beta$) E-16 (erg cm$^{-2}$ s$^{-1}$) &8.49     & & & & &2.16 & & & &  &18.8   \\
\hline
\multicolumn{7}{l}{$^a$ Dereddened line intensities from \citet{vanZeeSkillman2006}.}
\label{table:intensities}\\
\hline \\
\end{tabular}
\end{table}
\end{landscape}

\begin{table}
\begin{landscape}
\caption{Physical parameters, ionic and total abundances}
\begin{tabular}{lllrcccllllll}
\hline\hline
       &     \multicolumn{3}{c}{                            \ion{H}{II}  West}  &               \multicolumn{3}{c}{     \ion{H}{II} East  }   &
\multicolumn{3}{c}{ PN }   &\\
\cmidrule(lr){2-4}  \cmidrule(lr){5-7} \cmidrule(lr){8-10} 
Ionic ab  &       \multicolumn{2}{c}{ Direct Method}  &             Model&              \multicolumn{2}{c}{ Direct Method} &    Model  &  \multicolumn{2}{c}{ Direct Method} &    Model  \\
\cmidrule(lr){2-3} \cmidrule(lr){4-4} \cmidrule(lr){5-6} \cmidrule(lr){7-7} \cmidrule(lr){8-9}  \cmidrule(lr){10-10}
                      &                   value  &   $\sigma$ &    value &        value  &          $\sigma$ &       value &  value  &          $\sigma$ &       value  \\
                      \cmidrule(lr){2-3} \cmidrule(lr){4-4} \cmidrule(lr){5-6} \cmidrule(lr){7-7} \cmidrule(lr){8-9}  \cmidrule(lr){10-10}
n$_e$ (cm$^{-3}$)   		&113			& 121 		&   100 		& <100	&     ---	& 100 		& 1000* 	& --- 			& 1000\\
T[\ion{O}{III}]  (K)      		&17046 	& 2925 		& 14790 	& --- 		& --- 		&13994 	& 20897 	& 931 		& 18755 \\
T(\ion{O}{II})*  (K)     					&14932 	& 2048 		& 13889 	& --- 		& --- 		&12402 	&17628 	& 651 		& 17550\\
He$^+$/H$^+$(E-02)		& 9.45		& 1.20		& 7.12		&---		& ---		&0.102		&7.47		& 0.74		&	7.91	\\	
He$^{+2}$/H$^+$(E-02)	& ---			& ---			& ---			&---		& ---		& ---			&1.93		&0.14		&2.08\\
O$^+$/H$^+$(E-06)       &14.7	  	&2.02	  	& 21.1	  	& ---		& --- 		& 24.0 		& --- 			& --- 			& 0.43\\
O$^{+2}$/H$^+$(E-06)  & 9.89	  	&4.13    	&13.4  		& ---  	& --- 		& --- 			& 19.3 		& 0.86 		& 23.1 \\
N$^+$/H$^+$(E-06)       & 0.30	  	&0.04   		&0.29  		& --- 		& --- 		&0.30 		& 0.15 		& 0.01 		& 0.11\\
S$^+$/H$^+$(E-06)       & 0.23	  	&0.03   		&0.24  		& ---   	& ---  	& --- 			& --- 			& --- 			& 0.01\\
Ne$^{+2}$(E-06) 			& --- 			& --- 			& 0.95 		& ---		& --- 		& --- 			&3.03 		& 0.06 		& 4.59\\
Ar$^{+2}$(E-06)				& --- 			& --- 			& 0.19 		& --- 		& --- 		&0.002 		&0.026		& 0.006 	& 0.027\\
\hline
\multicolumn{7}{l}{* Adopted density. T(\ion{O}{ii}) from \citeauthor{CampbellTerlevich1986} expression, Eq. \ref{eq:Campbell}.} \\
\hline
       &        \multicolumn{5}{c}{   \ion{H}{II}  West}  &                          \multicolumn{1}{c}{    \ion{H}{II} East }   & \multicolumn{3}{c}{PN}\\
\cmidrule(lr){2-6} \cmidrule(lr){7-7} \cmidrule(lr){8-10} 
Total ab &    \multicolumn{2}{c}{ Direct Method}  &          \multicolumn{2}{c}{ONS}&   Model &   Model & \multicolumn{2}{c}{Direct method}& Model\\
\cmidrule(lr){2-3} \cmidrule(lr){4-5} \cmidrule(lr){6-6} \cmidrule(lr){ 7-7} \cmidrule(lr){ 8-9} \cmidrule(lr){10-10}
                      &                   value  & $\sigma$ &    value &     $\sigma$&    value  &                value & value & $\sigma$ &       value\\
                      \cmidrule(lr){2-3} \cmidrule(lr){4-5} \cmidrule(lr){6-6} \cmidrule(lr){ 7-7} \cmidrule(lr){ 8-9} \cmidrule(lr){10-10}        
12+log(He/H)  		& 10.98 	& 0.06           & --- 		& --- 			&11.00    	&  11.00 	& 10.97 	& 0.04 		& 11.00 \\
12+log(O/H)   		& 7.39  		&  0.10   		& 7.40  & 0.15  		&7.55	 	&7.40 		& 7.34 		& 0.02 	 	& 7.45\\
12+log(N/H)   		& 5.74 		&  0.09   		& 5.58  & 0.21  		&5.80	     &5.80  		& ---  		&---			& 7.00 \\
log(C/O)				& ---  		&  ---  			&---   	&  ---			& -0.80  	& -0.80 		& ---			& ---			& -1.0\\
log(Ne/O)				&--- 			& ---				& ---		& --- 			& -0.80 		& -0.80 		& -0.77		& 0.01		&-0.7\\
log(Ar/O)				&---   		& ---  			&   --- 	&  --- 		&  -2.10   	& -2.10 		&-2.65 		& 0.12			& -2.6\\
log(S/O)      			&-1.19 		& 0.04    		& ---    	& ---   	 	&  -1.40 	& -1.70 		& ---			&--- 			& -1.70\\
T(star) (E4 K)     		& ---			& ---   			& ---     	& ---			&  3.56	 	& 2.80		&--- 			&--- 			& 12.50\\
log(g) (star)   			& ---			& --- 				& --- 		& ---			&  3.90    	& 4.0 		&--- 			& --- 			& 7.0\\
<log U>       			& ---			& ---				& ---		& ---			& -2.19 		& -2.49 		&--- 			&--- 			& -2.12\\
log(Q(H$^{0}$))  	& ---			& ---				&--- 		& ---			& 48.9 		& 48.0 		&--- 			&--- 			& 48.0\\
r$_{in}$ (E17 cm)	& ---			& ---				& ---		&---			& 1.00  		&1.00  		& ---			& ---			& 1.00\\
r$_{out}$ (E18 cm)	&  8.97		& ---				& ---		& ---			& 9.63 		& 4.9 		&--- 			& ---			& 0.96\\
\hline \\
\label{table:abundances}
\end{tabular}
\end{landscape}
\end{table}
 
\end{document}